\documentstyle[aps,graphicx,tighten,preprint]{revtex}
\begin{document}
\title{Violations of Bell Inequalities for Measurements with 
Macroscopic Uncertainties: What does it Mean to Violate Macroscopic 
Local Realism? \\ }
 \vskip 1 truecm
\author{M. D. Reid\\ }
\address{Physics Department, University of Queensland, Brisbane, Australia\\ }
 \date{\today}
\maketitle
\vskip 1 truecm
\begin{abstract}
We suggest to test the premise of ``macroscopic local realism''  
which is sufficient to derive 
Bell inequalities when measurements of photon number 
are only accurate to an uncertainty of order $n$ photons, where $n$ is 
macroscopic. Macroscopic local realism is only 
sufficient to imply, in the context of the original 
Einstein-Podolsky-Rosen argument, fuzzy ``elements of 
reality'' which have a macroscopic indeterminacy. 
We show therefore how the violation of local realism in the presence of 
macroscopic uncertainties implies the failure of ``macroscopic local 
realism''.  
Quantum states violating this macroscopic local realism are presented. 
\end{abstract}
\narrowtext
\vskip 0.5 truecm

\section{Introduction}

There is increasing evidence for the failure of ``local realism'' as 
defined originally by Einstein, Podolsky and Rosen $^{\cite{1}}$, 
Bohm $^{\cite{2}}$ and Bell $^{\cite{3,4}}$. For certain correlated quantum
systems, Einstein, Podolsky and Rosen argued that 
``local realism'' is sufficient to 
imply that the results of measurements are predetermined. These 
predetermined ``hidden variables'' (called ``elements of reality'' by  
Einstein, Podolsky and Rosen) exist to describe the 
value of a physical quantity, whether or not the measurement 
is performed, and as such are not part of a quantum description. 
Bell later showed that the predictions of quantum mechanics 
for certain ideal quantum states could not be compatible with such hidden 
variable theories.
 
It is now widely accepted therefore, as a result of Bell's theorem 
and related experiments $^{\cite{5}}$, that local 
realism must be rejected. However the rejection of local realism implied 
by these results is at the most microscopic 
level of a single photon, in the sense that the ``hidden variables'' (or 
``elements of reality'') and the experimental results of measurements 
involved must be 
defined to the precision of one photon or better in order to prove local realism invalid.  
The failure of local realism in microscopic systems 
has long been associated with the existence of entangled quantum superposition 
states. In microscopic systems there can only be superpositions of 
states microscopically distinct. 

 Little is known about the validity of local realism 
in more ``macroscopic experiments'', where experimental uncertainties are 
larger (becoming macroscopic) in size, in an absolute sense. At 
least there has to our knowledge been no formulation of a Bell-type theorem or related 
experimental investigation for such a situation. Previous 
works $^{\cite{6,7,8}}$ 
suggestive of incompatibilities of local realism in macroscopic systems have 
considered the case where the measurements are performed with perfect 
accuracy and are thus not examples of macroscopic experiments as we have 
defined them here.

Generally it is thought that true macroscopic quantum effects come about 
from quantum superpositions of states which are macroscopically 
distinguishable $^{\cite{9,10}}$, often referred to 
as ``Schrodinger cat'' states. This point has been much discussed. 
Leggett and Garg $^{\cite{10}}$ have shown incompatibility of such 
macroscopic quantum states with the combined premises of ``macroscopic realism''  
and  ``macroscopic noninvasive measurability''. They however 
considered macroscopic quantum superposition states at a single location only, 
 and did not introduce the premise of locality. 

There has been much interest and debate 
over whether or not ``Schrodinger cat'' states can truly exist. 
The existence, for which there is now experimental evidence $^{\cite{11,12,13}}$,
 of ``Schrodinger cat'' states would appear to be 
closely linked to the question of the validity of local realism at the 
more macroscopic level we have described. One would suspect that a 
violation of local realism evident in an experiment where 
uncertainties are large would be due more to entangled macroscopic 
superpositions than microsuperpositions. 

 In this paper we begin (in Section 2) by defining 
 the physical premise of ``macroscopic local realism'' $^{\cite{14}}$ 
 so as to 
 identify the peculiar features of the macroscopically entangled 
 quantum 
 states in a way which is independent of the quantum formulation. 
 Macroscopic local realism is only sufficient to assign, to a system, 
 predetermined ``elements of reality'' (or hidden variables) 
 which are intrinsically 
 macroscopic, in that they have a macroscopic indeterminacy in their 
 values.
  
  Suppose 
our ``Schrodinger's cat'' is correlated with a second 
system spatially separated from the cat, for 
example a gun used to kill the cat. Let us suppose a
 gun that has been fired implies a dead cat; a gun that has not been 
fired impies a cat that is alive. We can predict the 
result for a measurement of the cat (whether dead or alive), 
without disturbing the cat,
 by measurement on the gun. Macroscopic local 
realism is the premise used to 
imply the existence of an element of reality for the cat. 
 The element of reality in this case is a variable 
 that assumes one of two values, one value corresponding to the ``dead'' 
 state and the other value corresponding to the ``alive'' state. The 
 assignment of this element of reality then means that the cat is 
 always either ``dead'' or ``alive'', regardless of whether or not 
 it is being observed or measured. Macroscopic local realism is used in 
this case, as opposed to local realism as used originally by Bell and 
Einstein et al, because the 
two possible results of measurement of the cat, ``alive'' and ``dead'', are 
macroscopically distinct. To summarise, the rejection of  
macroscopic local realism in this example means that 
we cannot think of the cat as being 
either dead or alive, even though we can predict the ``dead'' or ``alive'' 
result of ``measuring'' the cat, without disturbing the cat, 
by measuring the correlated spatially-separated second system, 
which in this case is the gun. The rejection of macroscopic local 
realism is a more startling result than, and is not implied by, the rejection of local 
realism as indicated by Bell's theorem.

In Section 3 we point out that 
 for situations where the possible  
results are all macroscopically  distinguishable, we need only assume 
the  strong premise of macroscopic local realism in order to derive the  
Bell inequalities. We then 
focus attention on the more general case where the results of measurement 
may be microscopically separated. We show that with the addition of 
macroscopic classical 
noise sources which model a macroscopically imprecise measurement, 
one may derive the Bell 
inequalities using only the premise of macroscopic local realism.  
Thus if a violation of a Bell inequality is maintained 
  in the presence 
 of macroscopic uncertainties in the measurement process, we have direct 
 evidence for an incompatibility with macroscopic local realism. 
 
  In Section 4 quantum states are presented which show a violation of Bell's 
inequality with such macroscopic noise, thus indicating an 
incompatibility of the predictions of quantum mechanics with the very 
strict form of macroscopic local 
realism we have defined. We believe this is the first such result 
although some preliminary results presented in this paper 
have been published previously $^{\cite{15}}$.
 Such a test of macroscopic local realism provides an avenue to 
 focus the peculiar macroscopic nonlocal aspects of the 
 ``macroscopic entangled quantum state''.

  The application of Bell inequality theorems, and the effect of noise  
on the violations predicted, to situations where many 
photons fall on a detector is relevant to the question of whether or not 
tests of local realism can be conducted in the experiments such as those 
performed by Smithey et al $^{\cite{16}}$. Here correlation of photon number between two 
spatially separated but very intense fields is sufficient to give 
``squeezed'' noise levels. In these high flux experiments detection losses 
can be relatively small, allowing for the possibility of violation of a 
strong Bell inequality, but noise which limits the resolution of the 
photon number measurement can be large in absolute terms, 
as compared to traditional Bell inequality 
experiments which involve photon counting with low incident photon numbers.

  \section{Definitions of Macroscopic Local Realism}

  In the original argument of 
 Einstein, Podolsky and Rosen $^{\cite{1}}$ (EPR), ``local realism'' is 
 defined in the following way. 
 ``Realism'' is sufficient to state 
 that if one can predict with 
 certainty the result of a measurement of a physical quantity 
 at $A$, without disturbing the 
 system $A$, then the results of the measurement were predetermined and 
 one has an 
 ``element of reality'', corresponding to this 
 physical quantity. The element of reality is a variable which assumes 
 one of the set of values which are the predicted results of the 
 measurement.  Locality postulates that measurements 
at $B$ cannot disturb $A$ in any way. 
Taken together with realism then, as defined above, 
 ``local realism'' is sufficient to 
 imply that, if one can predict the result of a measurement at $A$, by 
 making a simultaneous measurement at $B$, then the result of the 
 measurement at $A$ is a predetermined property of the system $A$.
 In the case of perfect correlation 
 and perfect measurements, the predetermined value (the element of reality) 
 for any individual system $A$ will have zero 
 uncertainty, since we can determine it precisely by measurements on $B$, 
 and because all orders of 
 change to system $A$ as a result of the measurement at $B$ are excluded 
 by locality.

 Macroscopic local realism may be defined 
   as a premise stating the following. 
 This meaning and definition of ``macroscopic local realism'' 
 has been previously introduced in references $[14]$ and [15] in line with 
 the original EPR 
   argument, and its experimental realisation for continuous 
   variables by Ou et al $^{\cite{17}}$. If one can predict 
   the result of a measurement at $A$ by performing 
 a simultaneous measurement on a spatially separated system $B$, then the 
 result of the measurement at $A$ is predetermined but 
 described by an element of reality 
 which has an indeterminacy in each of its possible  values, so 
 that only values macroscopically 
 different to those predicted are excluded. 
  We note that the meaning of 
 ``predict'' in the above definition could be loosened to allow an 
 uncertainty in the prediction, as one would have in macroscopic 
 experiments which incorporate measurement uncertainties.

Macroscopic local realism incorporates two assumptions. 
We 
define a ``macroscopic locality'', which states that measurements at a 
location $B$ cannot instantaneously induce changes of a macroscopic magnitude 
(for example the dead to alive state of a cat, or a change 
between macroscopically 
different photon numbers)  
 in a second system $A$ spatially 
separated from $B$. Locality in its entirety, as used 
originally by EPR and Bell, postulates that measurements 
at $B$ cannot disturb $A$ in any way.  We expect that our definition of a 
macroscopic order of locality is equivalent to 
postulating that locality will always be appear to be satisfied 
in experiments where measurement 
uncertainties do not enable resolution of results different by a 
microscopic or mesoscopic number of photons.

The second assumption incorporated by macroscopic local realism is the 
assumption of a ``macroscopic realism'',  since macroscopic local 
realism implies 
elements of reality with (up 
 to) a macroscopic indeterminacy. Suppose an element of reality may be symbolised by the variable $x$, 
where $x$ can take on numerical values $x_1, x_2, ...$. For  
microscopic realism these values are specified to a microscopic level. 
For ``macroscopic 
realism'' these values have a macroscopic indeterminacy, by this meaning 
 that one can only 
exclude values for the associated 
physical variable which are macroscopically different to the values $x_1, 
x_2,....$. We see that if $x_1, x_2$ are only microscopically distinct, 
they are in this case no longer distinguished by different hidden 
variable values.  

 The notion of 
realism is exclusive of 
``quantum superposition states'' in the following sense. If a physical 
quantity for an ensemble of systems is attributed an 
element of reality $x$ as 
above, then the element of reality for each individual system will take on 
one of the values $x_1, x_2, ...$. This value is the result of the 
measurement of the physical quantity, should it be performed. This 
element of reality picture is different to the standard 
quantum picture of a system being in a ``quantum 
superposition'' of two states of different $x_i$. According to standard quantum 
mechanics interpretation, an  individual system described by such a superposition  
cannot be thought of as being in one or other of the two states prior to 
measurement. If the values of the element  
of reality are defined with zero uncertainty then the element of 
reality picture excludes (or is different in its interpretation to) 
a ``quantum superposition'' of states $x_i$ and 
$x_i+\delta$ where $\delta$ is 
nonzero. 

We consider the existence of   
an element of reality which is only macroscopically specified, having  
values that can only be 
specified not to be macroscopically different to a value $x$. This 
macroscopic realism  
description says nothing 
about the possibility of superpositions of states microscopically or 
mesoscopically different to $x$. Macroscopic local realism cannot exclude the 
 possibility of ``quantum superpositions'' of states micro- or meso-scopically 
 different, with respect to the physical quantity represented by the 
 element of reality. We can however exclude the possibility 
of quantum superpositions of states with macroscopically different values 
for the physical quantity concerned.

Since it says nothing about microscopic systems, 
macroscopic local realism is a less 
 restrictive  
 premise than ``local realism'' used in its entirety. Local realism 
 in its full sense 
  can define elements of reality with values having no 
 uncertainty and therefore can exclude the 
 possibility of quantum superpositions of states with all separations 
 (micro to macro inclusive) in the relevant variable.

 \section{Bell Inequalities with Noise: Tests of 
  Macroscopic Local Realism }

   Our proposed experiment to test macroscopic local realism 
   is depicted in Figure 1, where 
    $\hat a_{\pm}$ and $\hat b_{\pm}$ are boson operators for outgoing 
    fields, generated from a suitable source to be discussed in 
    Section 4, 
    at the spatially separated locations $A$ and $B$ respectively.      
     We define the Schwinger spin operators
     \begin{eqnarray}   
   \hat S_{x}^{A}&=&(\hat a_+^\dagger \hat a_- + \hat a_-^\dagger 
   \hat a_+)/2 \nonumber \\
  \hat S_{y}^{A}&=&(\hat a_+^\dagger \hat a_- - \hat a_-^\dagger \hat a_+)/2i \nonumber \\
  \hat S_{z}^{A}&=&(\hat a_+^\dagger \hat a_+-\hat a_-^\dagger \hat a_-)/2.
  \end{eqnarray}
Similar operators 
  $\hat S_{x}^{B}, \hat S_{y}^{B},\hat S_{z}^{B}$ are defined for the modes 
  at $B$. We measure simultaneously at $A$ and $B$ the Schwinger spin operators
      \begin{eqnarray}
 \hat S_{\theta}^A&=& \hat S_{x}^{A}¥\cos{\theta} +\hat S_{y}^{A}¥\sin{\theta}
  \end{eqnarray}  
and
 \begin{eqnarray}
  \hat S_{\phi}^{B}&=& \hat S_{x}^{B}¥\cos{\phi} +\hat S_{y}^{B}¥\sin{\phi}
       \end{eqnarray}
    respectively.
  
In Figure 1a the measurement at $A$ is performed with phase shift 
$\theta$ and beam splitter to produce $\hat{c}_\pm^{'}¥=\left(\hat 
a_+\pm \hat a_- \exp (-i\theta) \right)/\sqrt{2}$, followed by 
photodetection. At $B$ modes $\hat{d}_\pm^{'}¥=\left(\hat 
b_+\pm \hat b_- \exp (-i\phi) \right)/\sqrt{2}$ are similarly generated.   
 The possible 
 outcomes for the photon number $\hat c_+^{'\dagger} \hat c_{+}¥^{'}¥$ (and 
 $\hat d_+^{'\dagger} 
 \hat d_+^{'}¥$) are 
 $0, 1, ...$ in integer steps. 
The spin values for $\hat S_{\theta}^A$ and  $\hat S_{\phi}^{B}$ 
 are then given by the photon number differences $
    \hat n_{\theta}^{A}=2\hat S_{\theta}^A=
\hat c_+^{'\dagger} \hat c_+^{'}¥-\hat c_-^{'\dagger} \hat c_-^{'}$ and 
$\hat n_{\phi}^{B}=2\hat S_{\phi}^B=
  \hat d_+^{'\dagger} \hat d_+^{'}¥-\hat d_-^{'\dagger} \hat d_-^{'}$. 

 Alternatively in Figure 1b, the  
 $\hat a_{\pm}$ are first combined $^{\cite{18}}$  
    through a beam splitter, and phase shifted, to give outgoing fields 
    $\hat a_{-}^{'}=(\hat a_{-}-\hat a_{+})/\sqrt{2}$ and 
    $\hat a_{+}^{'}=i(\hat a_{-}+\hat a_{+})/\sqrt{2}$. These may now be considered 
    system fields, upon which the measurement 
    $\hat n_{\theta}^{A}=2\hat S_{\theta}^A= 
    \hat c_+^\dagger \hat c_+-\hat c_-^\dagger \hat c_-$ 
    is made through the 
transformation (with polariser or beam 
    splitter) $\hat c_{+}=\hat a_{+}^{'}\cos{\theta/2}+ \hat a_{-}^{'}\sin{\theta/2}$ 
and $\hat c_{-}=\hat a_{+}^{'}\sin{\theta/2}- \hat a_{-}^{'}\cos{\theta/2}$ followed 
by photodetection.   
 Figure 1b depicts a measurement   
$\hat S_{z}^{A'}\cos{\theta} +\hat S_{y}^{A'}\sin{\theta}$ made on system 
operators $\hat a_{\pm}^{'}$, but is the same 
measurement depicted in 1a for the fields $\hat a_{\pm}$.  We use similar definitions  
  $\hat S_{x}^{A'},\hat S_{y}^{A'}$ and $\hat S_{z}^{A'}$ for the Schwinger operators in 
  terms of $\hat a_{\pm}^{'}$. Similar transformations are defined 
  for the measurement at $B$. We present this scheme because, for the 
  particular choice of quantum state discussed in Section 4, it 
  ensures both fields $\hat a_{\pm}^{'}$ incident on the measurement 
  apparatus (polarizer) can be macroscopic. This arrangement then is 
  crucial in providing a test of macroscopic local realism.

    We classify the result of our measurement as $+1$ if the result 
    for the photon number 
    difference measurement $\hat n_{\theta}^{A}$ or  
    $\hat n_{\phi}^{B}$
  is positive or zero, and $-1$ otherwise. The results at $B$ are 
  classified similarly.
  We build up the 
following probability distributions: $P_{+}^{A}(\theta)$ 
for obtaining $+$ at $A$; $P_{+}^{B}(\phi)$ for obtaining $+$ at $B$; and 
$P_{++}^{AB}(\theta,\phi)$ the joint probability of obtaining $+$ at 
both $A$ and $B$.

  We first 
  consider the predictions as given by the original definition of 
  local realism (local hidden variables) used by 
  Einstein-Podolsky-Rosen, Bell and 
  Clauser-Horne $^{\cite{3,4}}$ .
 The probability of obtaining $+1$ for $S_\theta ^A$ is expressed as  
  \begin{eqnarray}
P_{+}^A(\theta )=\int \rho(\lambda) \quad p_{+}^A(\theta, \lambda ) \quad 
d\lambda.  
\label{1}
\end{eqnarray}
The probability of obtaining `+1' for $S_\phi ^B$ is 
\begin{eqnarray}
P_{+}^B(\phi )=\int \rho(\lambda)\quad p_{+}^B(\phi, \lambda ) \quad d\lambda.  
\label{2}
\end{eqnarray}
The joint probability for obtaining `+1' for both of two
simultaneous measurements with $\theta$ at $A$ and $\phi$ at $B$ is 
\begin{eqnarray}
P_{++}^{AB}(\theta ,\phi )= \int \rho(\lambda) \quad p_{+}^A(\theta, \lambda ) 
p_{+}^B(\phi, \lambda )\quad d\lambda  
\label{3}
\end{eqnarray}
Here $p_{+}^A(\theta, \lambda )$ is the probability   
  for getting the result $+1$ given the hidden variables 
  $\lambda$; $p_{+}^B(\phi; \lambda )$ is the probability   
  for getting the result $+1$ given  
  $\lambda$; while $\rho (\lambda)$ is the probability distribution 
  for the hidden variables $\lambda$.
    
It is well known $^{\cite{3,4}}$ that one can  derive the following
 ``strong'' Bell-Clauser-Horne inequality from the assumptions of local 
 realism made so far. 
  \begin{eqnarray}
 S={{P_{++}^{AB}(\theta,\phi)-P_{++}^{AB}(\theta,\phi')+P_{++}^{AB}(\theta',\phi)
 +P_{++}^{AB}(\theta',\phi')}\over{P_{+}^{A}(\theta')+P_{+}^{B}(\phi)}} \leq 1
	\label{4}
\end{eqnarray}
 To date this ``strong'' inequality  has 
  not been violated in any experiment, because of the poor detection 
  inefficiencies which occur in photon counting experiments. 
   It is well documented that it is possible to derive, with the 
  assumption of additional premises, a weaker form of the Bell inequality 
  which has been violated in photon counting experiments where detection 
  losses are high. In this paper however we restrict attention to the 
  strong inequalities which do not require additional assumptions. 
  Our proposed experiments involve photodiode detectors which have 
  high efficiencies and therefore allow the possibility of a strong 
  violation of local realism.
     
  In deriving the Bell inequalities, one specifies a probability 
  $p_{+}^A(\theta, \lambda )$  
  for getting the result $+1$ as opposed to $-1$ given the hidden variables 
  $\lambda$. If the results $+1$ and $-1$ are always macroscopically 
  different, it becomes apparent that one need only assume ``macroscopic 
  local realism'' as opposed to local realism in its entirety to obtain 
  the Bell inequalities. This is because in assuming the independence of 
  this probability $p_{+}^A(\theta, \lambda )$  on $\phi$, we 
  need only assume a macroscopic locality, 
  that the measurement at $B$ does not disturb the system at $A$ in a 
  macroscopic way to make the change from $+1$ to $-1$.  
  The elements 
  of reality need only be specified ``macroscopically'', that is they can 
  have a macroscopic indeterminacy in their values, and still adequately 
  represent the distinct outcomes of measurement. We can add certain 
  (though not all) perturbations of a macroscopic size (in photon number) 
  to the values predicted by the ``elements of reality'' and not change 
  the final form of the Bell inequality.

  The violation of the Bell inequality (7), where the possible results 
  of all relevant measurements 
  (for all relevant angles $\theta$ and $\phi$) are macroscopically 
  distinct, would be firm confirmation of an incompatibility with 
  macroscopic local realism. To our knowledge no such violation has yet been 
  demonstrated.

  In order to test conclusively for macro forms of local realism 
  in more general situations (where the possible results are not always 
  macroscopically separated), we propose to   
 add local noise sources to the final readout stage of each of the measurement processes, at $A$ and 
 $B$. We will assume that the result for the photon number difference 
 $\hat n_{\theta}^{A}$ or $\hat n_{\phi}^{B}$
    at each of 
   $A$ and $B$ respectively is of the form  
   $ n + noise$, where $n$ is the result of the measurement
    in the absence of the noise and 
  $noise$ is a local classical noise term. 
  The noise terms 
  at $A$ and $B$ are  independent, modeling a local physical source of 
  noise, and as such always satisfy locality, the noise added at $A$ 
  for example being independent of the experimental choice of the angle $\phi$ at $B$.
  
   We will derive a Bell inequality based on the premise of macroscopic 
  local realism alone, by showing that the addition of this classical noise 
  to the final measurement result can   
  alter the premises needed to derive the Bell inequality.

 We first define the  
 probability $P_{ij}^{0,AB}(\theta,\phi)$ for 
 obtaining results $i/2$ and 
 $j/2$ respectively upon joint measurement of 
 $S_{\theta}^A$ at $A$, and $S_{\phi}^B$ at $B$, 
  in the absence of the applied noise. The $i$ and $j$ are then 
  results for the photon number differences 
  $\hat n_{\theta}^{A}$ or $\hat n_{\phi}^{B}$ respectively. 
 In terms of a local  
  hidden variable description, 
  this probability is given by
    \begin{eqnarray}
   P_{ij}^{0,AB}(\theta,\phi) = \int \rho(\lambda) \quad p_{i}^A(\theta, \lambda ) 
   p_{j}^B(\phi, \lambda )\quad d\lambda  
\end{eqnarray}

  We next outline how the assumption 
 of local realism, as defined originally by EPR, implies the hidden 
 variable description (8) above. This is in order to postulate how the 
 above expression is modified if one makes only the macroscopic local 
 realism assumption.

A perfect correlation between  
measurement results at $A$ and $B$ is predicted possible for some 
quantum states.   For such situations, it is possible to predict 
precisely the 
result of a measurement at $A$ by performing a particular 
measurement at $B$.
We are able to deduce $^{\cite{3}}$, assuming 
  local realism and following the reasoning of EPR as outlined in Section 2, 
  the existence of a  set of  
  ``elements of reality'', $m_\theta ^A$ and $m_\phi ^B$, one for each subsystem at $A$ and $B$, and one 
  for each choice of measurement angle, $\theta$ or $\phi$, at $A$ or $B$ 
  respectively.  The  $m_{\theta}^{A}$
assumes one of a set of definite values, this value giving the 
result of the measurement $\theta$ at $A$ should it be performed. 
The set {$m_{\theta}^{A}, m_{\phi}^{B}$} forms a set of hidden variables 
$\lambda$ for the system.

  More generally there will be a reduced 
  correlation 
  between measurements performed at $A$ and $B$. 
  This is generally so for the case where measurements incorporate macroscopic 
uncertainties. 
 Local 
  realism still allows us to deduce the existence of an element 
  of reality (we will call it $m_{\theta}^A$) for the photon number 
  difference at $A$, with 
  measurement angle $\theta$ at $A$, since we can make a prediction of the 
  result at $A$, without disturbing the system at $A$, under the locality 
   assumption. This prediction is based on a measurement performed 
   at $B$. In this case however the element of reality $m_{\theta}^{A}$ becomes 
``fuzzy''. The ``values'' which the element 
  of reality can assume do not form a set of definite 
  numbers with zero uncertainty, but rather a set of distributions, one for each 
  possible result $m$ at $B$, which we label by $m_{\theta}^A=m$. The 
  distribution labeled by the element of reality  $m_{\theta}^A$ assuming the value $m$ 
  gives the probability of a result for the measurement 
 $\theta$ at $A$  
 should it be performed. It is independent of $\phi$ the 
 experimenter's choice of angle at $B$ if a simultaneous measurement 
 at $B$ should be performed. 
   One can apply similar 
  reasoning to deduce the existence of a set of indeterminate elements of reality 
  $m_{\phi}^B$.

 The assumption of ``local realism'' then justifies the local hidden 
 variable description used in (8), and (4)-(6), above. 
 Local realism implies that the 
 system is always in a state corresponding to a particular 
 value for each of the elements of reality ${m_{\theta}^A}$ and 
 ${m_{\phi}^B}$. 
  The whole set of ``elements of reality'' 
${m_{\theta}^A}$ and ${m_{\phi}^B}$ form a set of ``hidden variables'' 
which can be attributed to the system at a given time.   
Common notation symbolises the complete set of hidden variables by 
$\lambda$, and the underlying joint probability distribution 
$p ({m_\theta ^A}, {m_\phi ^B})$  becomes $\rho (\lambda)$.
 The probabilities 
$\rho (\lambda)$ for the hidden variables are 
predetermined, and do not depend on the experimental choice of $\theta$ 
and $\phi$. 
For each such state $\lambda$ there is a 
probability $p_n^A (\theta, \lambda)$  
 that the result of a $\theta$ 
measurement at $A$ will be $n$. In the case with perfect 
correlation the ``elements of reality'' give precise 
values for the result of the photon number measurement. Suppose the 
result $m$ at $B$ correlates with $n$ at $A$.  Then we have   
$p_n^A (\theta, \lambda)=1$ if 
$\lambda=m_{\theta}^A=m$, and is zero 
otherwise. More generally we have imperfect correlation and 
``fuzzy'' elements of reality, meaning 
that this $p_n^A (\theta, \lambda)$ assumes a finite variance as 
discussed above.

 We focus attention on the distribution
  $p_i^A (\theta, \lambda)$, the probability of getting a photon 
  number $i$ for measurement at $A$ with angle $\theta$, 
  given that the system is in a hidden variable state $\lambda$. 
  The independence of  $p_i^A (\theta, \lambda)$ 
  on $\phi$ is based on the locality assumption used in its entirety, 
  that the experimenter's choice of measurement angle at $B$ cannot 
  (instantaneously) change the result of the measurement at $A$ in 
  any way. 
  With macroscopic local realism the locality condition is relaxed, 
  allowing the conditional 
  distributions $p_i^A (\theta, \lambda)$ to
   become nonlocal, that is to have an explicit 
  dependence on the experimental angle $\phi$. The locality condition 
  is relaxed however only up to the level of 
  $M$ photons, where $M$ is not macroscopic, by maintaining 
  that the measurement at $B$ cannot instantaneously 
  change the result at $A$ by an amount exceeding $M$ photons.

  By relaxing 
  the locality assumption up to $M$ photons, the elements of reality 
  $m_{\theta}^A$ (deduced by way of the EPR argument) even in 
   situations of perfect correlation will automatically have a
    distribution $p_i^A ({\theta,\phi}, \lambda)$  
   which is no longer a delta function, though 
   the distribution will be zero for values of $i$ exceeding the value of 
   $m_{\theta}^A$ by greater than $M$ photons. This is because we can no longer 
   exclude the possibility of changes to the result of photon number 
   measurements at $A$ by an amount of up to $M$ photons, 
   due to the measurement at $B$. 
   
   Similarly in the case of imperfect 
   correlation the ``fuzziness'' of the elements 
   of reality as given by the conditional distribution 
   $p_{i}^A(\theta, \lambda )$ is  increased, by an amount whose upper 
   limit is determined by 
   the value of $M$ and which may depend on $\phi$.

 Now we must consider the prediction for 
 equation (8) as given by macroscopic local realism.
    The elements of reality deduced using macroscopic local realism cannot give 
  predictions for the results of measurement  
  which are macroscopically different to those predicted from the 
  elements of reality deduced using local realism. Where our predicted result 
  for a measurement at $A$ is 
  $i'$ using local realism, macroscopic local realism 
  allows the result to be $i'+m_{A}¥$ where $m_{A}¥$ can be any 
  number not macroscopic. Importantly, while $i'$ is not dependent on 
  the choice $\phi$ for a simultaneous measurement 
  at $B$, the value $m_{A}¥$ can be. 
We therefore introduce the macroscopic locality assumption 
  into the expression (8) 
   for the probabilities in terms of the hidden variables in the 
   following manner. We assume that the conditional probability  
   $p_{i}^A(\theta, \lambda )$ in equation (8) takes the  
   form of the following convolution (where $M$ is a integer which 
   is not macroscopic). 
   
     \begin{equation}
     p_{i}^A({\theta,\phi}, \lambda )= \sum_{m_A=-M}^{+M}
      p_{m_A}^{A,NL} ({\theta,\phi},\lambda) 
      p_{i'=i-m_A}^{A,L} (\theta, \lambda ) 
\end{equation}
   (We similarly relax the 
  locality assumption for  $p_i^B (\phi, \lambda)$, allowing for a 
  dependence on $\theta$, and introduce a $p_i^B (\phi, \theta, \lambda)$ 
  defined in a similar fashion.)
   The original local probability distribution 
   $p_{i'}^{A,L} (\theta, \lambda )$, as would be specified 
   through local realism, may be convolved with a microscopic 
   or mesoscopic nonlocal probability function 
   $p_{m_A}^{A,NL} ({\theta,\phi},\lambda) $. The local 
   specification, which is not dependent on the experimental choice of angle 
   $\phi$ at $B$, gives a (local) probability distribution 
    $p_{i'}^{A,L} (\theta, \lambda )$ for obtaining 
   $i'$ photons at $A$, but the prediction is only correct to within $\pm M$ 
   photons. These (local) distributions form 
   the fuzzy ``macroscopic elements of reality''.
   The probability distribution for an actual result $i=i'+m_{A}¥$
    at $A$ is determined by the further nonlocal 
   perturbation term $p_{m_A}^{A,NL} ({\theta,\phi},\lambda)$, which gives 
   the probability of a further change of $m_A$ photons. 
   The nonlocal term is necessary because macroscopic local realism allows for   
    the possibility that the measurement at $B$ instantaneously changes the result 
   at $A$ by $M$ or less photons, where $M$ is not macroscopic. 
   The only restriction is that the nonlocal distribution does not 
provide macroscopic perturbations, so that the 
	probability of getting a nonlocal change outside the range $m_A=-M,...,
	+M$ is zero. 
	Equivalently we must have (and similarly for terms with $B$)  
\begin{equation}	
	 \sum_{m_A=-M}^{M}
       p_{m_A}^{A,NL} (i^{'},{\theta,\phi},\lambda)  = 1.   
  \end{equation}

  We now wish to obtain an expression for the 
  measurable probabilities $P_{++}^{AB}(\theta,\phi)$ 
  in the presence of the local noise terms, in terms of 
  the $P_{ij}^{0,AB}(\theta,\phi)$. We introduce noise distribution functions at 
  each of $A$ and $B$, and define probabilities such as  
  $P^A(noise\geq x)$, that the $noise$ at $A$ 
   is greater than or equal to the 
  value $x$. A probability $P^B(noise\geq x)$ is defined similarly, for 
  the noise term at $B$. The final measured probability in the presence of 
  noise is expressible as
      \begin{equation}
   P_{++}^{AB}(\theta,\phi) = \sum_{i,j=-\infty}^{\infty}
   P_{ij}^{0,AB}(\theta,\phi) P^A(noise\geq -i) P^B(noise\geq -j)
\end{equation}
 We write the predictions for this expression in terms of the 
hidden variable theory by substituting the macroscopic locality 
assumption (9) into the hidden variable prediction (8) for  
$P_{ij}^{0,AB}(\theta,\phi)$. We get
    \begin{eqnarray}
   P_{++}^{AB}(\theta,\phi) &=& 
   \sum_{i,j=-\infty}^{\infty} \int \rho(\lambda)\biggl[ \sum_{m_A=-M}^{M}
       p_{m_A}^{A,NL} (i^{'},{\theta,\phi},\lambda) 
      p_{i'=i-m_A}^{A,L} (\theta, \lambda )   \biggr.\nonumber \\
&\times&  \biggl.\sum_{m_B=-M}^{M} p_{m_B}^{B,NL} (j^{'},{\phi,\theta},\lambda) 
      p_{j'=j-m_B}^{B,L} (\phi, \lambda )\biggr] d\lambda \quad 
       P^A(noise\geq -i) P^B(noise\geq -j) 
\end{eqnarray}
Recalling $i=i'+m_A$ and $j=j'+m_B$ we change the $i$, $j$ summation to 
one over $i'$, $j'$ to get
  \begin{eqnarray}
   P_{++}^{AB}(\theta,\phi) &=& 
   \sum_{i',j'=-\infty}^{\infty}\int \rho(\lambda)p_{i'}^{A,L} (\theta,\lambda)
    \biggl[\sum_{m_A=-M}^{M}
        p_{m_A}^{A,NL} (i^{'},{\theta,\phi},\lambda) 
       P^A(noise\geq -(i'+m_A)) \biggr]  \biggr. \nonumber \\
 &\times&   p_{j'}^{B,L} (\phi, \lambda )\biggl[\sum_{m_B=-M}^{M}
\biggl.     p_{m_B}^{B,NL} (j^{'},{\phi,\theta},\lambda) 
     P^B(noise\geq -(j'+m_B)) \biggr] d\lambda  
\end{eqnarray}

 At this point we introduce the following assumption regarding the 
 macroscopic nature of the noise term  $P^A(noise\geq x)$, 
 that the increase or decrease of $x$ 
 by an amount of up to $M$  
 photons gives only a negligible change to the probability that the noise 
 is of size $x$ or greater, 
$ P^A(noise\geq -(i'+m_A)) \approx P^A(noise\geq -i') 
$ and similarly for the noise term at B. This gives us
 \begin{equation}
\sum_{m_A=-M}^{M}
        p_{m_A}^{A,NL} (i^{'},{\theta,\phi},\lambda) 
       P^A(noise\geq -(i'+m_A)) \approx 
        P^A(noise\geq -i') \sum_{m_A=-M}^{M}
        p_{m_A}^{A,NL} (i^{'},{\theta,\phi},\lambda).
   \end{equation}      
Clearly this is only valid for noise which is macroscopic in size 
(recalling $M$ is a number which is not macroscopic). With 
	assumption (10) we get simplification to obtain a final form  
	  \begin{equation}
   P_{++}^{AB}(\theta,\phi)=\sum_{i',j'}\ \int \rho(\lambda)
     p_{i'}^{A,L} (\theta, \lambda )  p_{j'}^{B,L} (\phi, \lambda ) d\lambda 
    \times P^A(noise\geq -i') P^B(noise\geq -j').
 	\end{equation}
  This prediction of the hidden variable 
	theory is now given in a (local) form like that of (6). Similar study 
	of the expressions for the marginal probabilities lead to (local) expressions like 
	that of (4) and (5), and the Bell 
	inequalities (7) therefore readily follow. The noise terms 
	$noise$ which add a macroscopic 
	uncertainty to the photon number result alter the premises 
	needed to derive the Bell inequality. We need only assume 
	macroscopic local realism to derive the 
	inequalities (7) in the presence of macroscopic noise terms.   Violation 
	therefore of 
	these Bell inequalities in the presence of the macroscopic noise terms 
	 would be evidence of a failure of macroscopic local 
	realism.
	
 \section{Quantum States Violating Bell Inequalities with Macroscopic 
 Noise: Predicted Failure of 
  Macroscopic Local Realism }  
 
We present a quantum state which shows violations of Bell 
inequalities in the presence of macroscopic noise. By the above 
arguments, this state then  
is evidence of a failure of macroscopic local realism.  
\begin{equation}
	|\psi> = [I_{0}(2r_{0}^{2})]^{-1/2}\sum_{n=0}^{\infty}
	\frac{(r_{0}^{2})^{n}}{n!}|{n}>_{a_{-}} |{n}>_{b_{-}} 
	|\alpha>_{a_{+}} |\beta>_{b_{+}}
\end{equation}
Here $I_{0}$ is a modified Bessel function.  
 The fields $\hat a_{+}$ and $\hat b_{+}$ are in coherent states  
$|\alpha>_{a_{+}¥}$ and  $|\beta>_{b_{+}¥}$ respectively 
 and we 
allow $\alpha$, $\beta$ to be real and large. $|{n}>_{k}$ is a Fock 
state for field $k$. The fields $\hat 
a_-$ 
and $\hat b_-$, often referred to as signal and idler fields 
respectively, are microscopic and are generated 
in a pair-coherent state  with $r_{0}=1.1$. 
Pair-coherent states were considered originally by Agarwal $^{\cite{19}}$. They 
might potentially be generated using  
nondegenerate parametric 
oscillation (as suggested by Krippner and Reid $^{\cite{19}}$ 
and explored in the recent work by Gilchrist and Munro $^{\cite{19}}$) in a 
limit where one-photon losses are negligible, or 
some similar process, as modelled by the following Hamiltonian 
in which coupled two-photon signal-idler loss 
dominates over linear single-photon loss:  
\begin{equation}
   H = i\hbar E ( \hat a_-^\dagger \hat b_-^\dagger - \hat a_- \hat b_-) + 
   \hat a_- \hat b_- \hat \Gamma ^\dagger + 
   \hat a_-^\dagger \hat b_-^\dagger \hat \Gamma
\end{equation}
The coherent states for $\hat a_{+}¥$ and $\hat b_{+}¥$ would be derived from 
the laser pump for the oscillator. 
Here E represents a coherent driving parametric term which generates signal-idler pairs, while 
$\hat \Gamma$ represents reservoir systems which give rise to the coupled 
signal-idler loss. The Hamiltonian preserves the 
signal-idler photon number difference operator 
$\hat a_-^\dagger \hat a_-$-$\hat b_-^\dagger \hat b_-$, 
of which the quantum state (16) is an eigenstate, with eigenvalue zero.
We note the analogy here to the single mode ``even'' and ``odd'' coherent 
superposition states $N_{\pm}^{1/2}(|\alpha> \pm |-\alpha>)$ (where 
$\alpha$ is real and $N_{\pm}^{-1}=2(1 \pm exp(-2|\alpha|^2)$) which are 
generated by the degenerate form (put $\hat a_-=\hat b_-$) 
of the Hamiltonian (17). These states for large $\alpha$ are analogous to the famous 
``Schrodinger-cat'' states $^{\cite{9,10}}$ and have been 
recently experimentally explored 
$^{\cite{11,12,13}}$. We point out later other choices of $|\psi\rangle$ 
possible.

To model noise we allow
 $noise$ to be a random noise term with a gaussian distribution of standard 
 deviation $\sigma$.
  An example of a noisy photon number measurement is photodiode 
  detection of very large intensities, 
  such as used in the experiments of Smithey et al $^{\cite{16}}$. 
 The photocurrent is processed electronically in 
  a way that adds noise to the final output current, giving a final 
  imprecision in the photon number measurement. 
   Although percentage detection efficiencies 
  are high for diode detectors, detection inefficiencies can also create a 
  potentially large absolute noise term which also limits the resolution of 
  the photon number measurement.

 Violations of the Bell inequality (7), for the state (16), in the absence 
 of noise are shown 
 in Figure 2, curve (a). The effect of adding increasing noise is to 
 reduce the value of 
 $S$ until eventually the violation is lost, at a cut-off noise value 
 $\sigma_{c}$, as shown in Figure 3.
 Figure 2, curve (b) shows this cut-off value $\sigma_{c}$ 
 (the maximum noise still allowing a violation of the Bell 
 inequality) versus $\alpha$. 
 We note the linear dependence of $\sigma_{c}$
  on $\alpha$ ($\sigma_{c}=.26\alpha$). In the 
 limit of larger $\alpha$ this cut-off noise $\sigma_{c}$ then 
 becomes macroscopic. 
 Violations of fixed magnitude ($S\rightarrow 1.0157$ as $\alpha 
 \rightarrow \infty$) are still  
 possible for increasingly larger absolute noise, simply by 
 increasing $\alpha$.

The asymptotic behavior in the large $\alpha$,$\beta$ limit is  
crucial in determining whether macroscopic local realism will be 
violated, and is understood by replacing the boson 
operators $\hat a_+$ and $\hat b_+$ by classical amplitudes $\alpha$ and 
$\beta$ 
respectively. We see that $\hat S_{\theta}^A$ from equation (2) can be 
expressed as  
$\hat S_{\theta}^A=(\hat a_+^\dagger  \hat a_- \exp (-i\theta)+\hat a_+  \hat 
a_-^\dagger \exp 
(i\theta))/2=\alpha \hat X_{\theta}^A/2$, and similarly   
$\hat S_{\phi}^B=\beta \hat X_{\phi}^{B}/2$, 
where  $\hat X_{\theta}^A=\hat a_-exp(-i\theta)+\hat a_-^\dagger exp(i\theta)$ 
and $\hat X_{\phi}^B=\hat b_-exp(-i\phi)+\hat b_-^\dagger exp(i\phi)$. 
The $\hat X_{\theta}^A$ and $\hat X_{\phi}^B$ are the quadrature phase 
amplitudes of the fields $\hat a_-$ and $\hat b_-$ respectively. 
We see then that the photon number measurements $2\hat S_{\theta}^A$ and  
$2\hat S_{\phi}^B$ give results in the
 large $\alpha, \beta$ limit corresponding numerically to the     
scaled quadrature phase amplitudes $\alpha \hat X_{\theta}^A$ and $\beta 
\hat X_{\phi}^B$ 
respectively. 
Figure 1a in fact shows for large $\alpha, \beta$ 
the experimental arrangement for balanced 
homodyne detection $^{\cite{20}}$, a
 technique commonly used to measure quadrature phase amplitudes. 
In Figure 1a the homodyne scheme measures 
the quadrature phase amplitudes $\hat X_{\theta}^A$ 
and $\hat X_{\phi}^B$, of the fields $\hat a_{-}¥$ and $\hat b_{-}¥$. The large 
intensity fields $\hat a_{+}¥$ and $\hat b_{+}¥$ are the ``local oscillator'' 
fields usually considered to be classical amplitudes $\alpha,\beta$.
 Violations of Bell inequalities (7) (failure of local realism) for precisely 
these asymptotic quadrature phase amplitude measurements have recently been shown by Gilchrist et al 
$^{\cite{21}}$, the value of 
$S=1.0157$ 
presented in these quadrature phase amplitude calculations indeed 
corresponding to our large $\alpha$ limit (Figure 2).
 
Calculations $^{\cite{22}}$ which model the addition of noise 
to the quadrature phase amplitude 
measurements $\hat X_{\theta}^A, \hat X_{\phi}^B$ reveal violations of the 
Bell inequality to be 
lost at the cutoff value of $\sigma_{0}¥=0.26$. This asymptotic 
result allows us to 
make a prediction of the effect of noise (in the large $\alpha$ limit) 
on the full photon number 
calculation presented in Figure 2.
The detected photon number difference is given as
\begin{equation}
\hat n_{\theta}^{A}=2\hat S_{\theta}^A=\hat{c}_{+}^{\dagger}¥\hat{c}_{+}¥
-\hat{c}_{-}^{\dagger}
\hat{c}_{-}¥= \alpha  \hat X_{\theta}^{A}¥
\end{equation}
 Noise of size $noise$ added to the photon number difference 
 $\hat n_{\theta}^{A}$ result is equivalent to noise of size 
 $noise/(\alpha)$ added 
 to the signal quadrature phase amplitude $\hat X_{\theta}$ result. The noise in 
 the photon number difference is scaled by a factor of $\alpha$, the local 
 oscillator amplitude. Therefore  the cut-off value $\sigma_{0}¥=0.26$ will 
correspond to a cut-off noise value of $\sigma_{c}=\alpha \sigma_{0}¥$ in the measurement of 
photon number difference $\hat n_{\theta}^A=2\hat S_{\theta}^A$, confirming the linear 
behavior shown in Figure 2, and the prediction that from this that 
macroscopic noise values are possible while still obtaining a 
contradiction with local realism. This property then is a predicted 
contradiction of quantum mechanics with macroscopic local realism as 
we have defined it.

The tolerance of the bell inequality violations to increasing noise as  
$\alpha$ increases can be understood as follows. For large $\alpha$, 
the shape (envelope) of the joint probability 
$P_{i,j}^{0,AB}(\theta,\phi)$, for obtaining results $i$ and $j$ upon 
measurements of $\hat{n}_{\theta}^{A}¥$ and $\hat{n}_{\phi}^{B}¥$ 
respectively, is determined by $P_{x,y}(\theta,\phi)$ (where 
$x=i/\alpha$ and $y=j/\beta$), the joint probability distribution for 
results $x$ and $y$ upon measurements $\hat{X}_{\theta}^{A}¥$ and 
$\hat{X}_{\phi}^{B}¥$  respectively. As the scale factor $\alpha$ 
linking result $x$ to result $i$ increases, the probability for 
obtaining a result in a region between two fixed, yet macroscopically- 
distinct photon numbers, will become small. In this limit we have 
macroscopically distinct outcomes, and measurements can tolerate a 
macroscopic noise without loss of violation of the Bell inequality.  
 
 Detection inefficiencies will also contribute to a noise in the final  
 result for the measurement, though in this case the noise will not be 
 gaussian. Noise caused by detector losses 
is often modeled by a  
beam splitter interaction immediately prior to photodetection. 
The field to be detected, $\hat c_{+}^{'}¥$ say, 
is taken to be an input to a beam 
splitter. The second input to the beam splitter $\hat{a}_{vac+}$ 
is considered to be a vacuum. The output 
\begin{eqnarray}
	\hat{c}_{L+}^{'}¥&=&\sqrt{\eta}\hat c_{+}^{'}+\sqrt{1-\eta}\hat{a}_{vac+},
	\label{eqn:loss}
\end{eqnarray}
where $\eta$ is the overall efficiency factor, is then 
taken to be the effective detected field. A similar effective field 
$\hat c_{L-}^{'}¥$ is constructed for the second detector, used to measure 
$\hat c_{-}^{'}¥$, at 
location $A$ and a second vacuum input $\hat{a}_{vac-}$ defined. 
The detected photon number difference is now given as
\begin{eqnarray}
\hat n_{\theta}^{A}=\hat{c}_{L+}^{'\dagger}\hat{c}_{L+}^{'}¥
-\hat{c}_{L-}^{'\dagger}\hat{c}_{L-}^{'}¥=
\eta \alpha  \hat X_{L\theta}^{A}¥
\end{eqnarray}
 where
 \begin{eqnarray}
 \hat X_{L\theta}^{A}¥=\eta \hat X_{\theta}^{A}¥+(\sqrt{1-\eta}/\sqrt{2})
 (\hat X_{\theta,vac+}+\hat X_{\theta,vac-})
\end{eqnarray}
 and the terms $\hat X_{\theta,vac\pm}$ are quadrature phase amplitudes 
 for the independent $+$ and $-$ vacuum modes representing the input 
 fields $\hat{a}_{vac+}$ and $\hat{a}_{vac-}$ respectively. Additional terms which give 
 negligible contributions with large $\alpha$ have been omitted. 
 We see how loss (described by $\eta$ less than 1) 
 causes a noise term $(\sqrt{1-\eta}/\sqrt{2}$)
 ($X_{\theta,vac+}+X_{\theta,vac-}$) in the signal quadrature phase 
 amplitude. Because of the factor $\eta \alpha$ this term can be 
 large to give 
 potentially macroscopic absolute noise values in photon number for the photon 
 number difference measurement. Violations of the Bell inequality 
 considered by Gilchrist et al $^{\cite{21}}$ have been shown obtainable in the presence of 
 detector losses ($\eta \approx .98$). We see from the above analysis that 
 this will 
 correspond for sufficiently 
 large $\alpha$ to a macroscopic absolute noise term 
 in the photon number measurements. Thus we have a second situation 
 where violations of a Bell inequality are predicted possible in the presence 
 of large absolute detector noise, this prediction indicating an 
 incompatibility of quantum mechanics with macroscopic local realism.
 
 We can deduce from our asymptotic (large $\alpha,\beta$) study  
 other states $|\psi\rangle$ which will give a failure of macroscopic local 
 realism. Any state $|\psi\rangle$ which shows a failure of 
local realism for 
measurements  $\hat X_{\theta}^A$ 
and $\hat X_{\phi}^B$ on fields $\hat a_{-}¥$ and $\hat b_{-}¥$ will also show a 
violation of macroscopic local realism, provided $\alpha$,$\beta$ are 
large. This follows because there will always be a finite noise cutoff 
$\sigma_{0}¥$, meaning that a failure of local realism is possible for 
noise values less than $\sigma_{0}¥$. For large enough $\alpha, \beta$ 
this cut-off will 
correspond to a macroscopic noise cut-off value $\sigma_{c}=\alpha \sigma_{0}¥$ 
in the photon number 
measurement $\hat n_{\theta}^A$ (and similarly for measurement 
$\hat n_{\phi}^B$). This is an important point since other states 
violating local realism for quadrature phase amplitude measurements, 
either by way of a Bell inequality or by way 
of the 
Greenberger-Horne-Zeilinger phenomenon, have been recently 
predicted $^{\cite{21}}$. This greatly increases the scope 
for a practical violation of 
macroscopic local realism.

A failure of local realism in the presence of 
macroscopic noise terms (as we have predicted here for states showing failure of 
local realism for quadrature phase amplitude measurements) is not typical. 
Consider as a source for the 
outgoing fields $\hat a_{\pm}^{'}$, pictured in Figure 1b, the following 
higher spin state which has been studied in much detail by Mermin 
and Drummond and others $^{\cite{6,7,8}}$. It is well known that this state gives a 
violation of Bell inequalities for large $N$, and is often considered 
to be an example of a violation of a ``macroscopic local realism''. 
\begin{equation} 
|\varphi \rangle = {{1}\over{ N! \left(N+1\right)^{1/2}}}
\left(\hat a_+^{'\dagger} \hat b_+^{'\dagger}+\hat a_-^{'\dagger} 
\hat b_-^{'\dagger}\right)^N |0\rangle
 |0\rangle
 \end{equation} 
Yet a study of the behaviour of the violation of the Bell inequality 
(7) with respect to noise added to the final photon number 
measurements gives a cut-off noise 
limit which is microscopic for large incident photon number $N$. This 
effect is plotted in Figure 4. This is in contrast with our state 
(16) which gives a macroscopic cut-off noise value in the limit of large $\alpha$.

It may be asked how a macroscopic claim can be made from the 
predictions discussed in this paper, given that the signal field 
$\hat a_{-}¥$ is 
microscopic.   
 It is noted in response to this question that, although the field 
 $\hat a_{-}¥$ is itself 
microscopic, the physical quantity measured, and to 
which the elements of reality relate, is 
the combined Schwinger operator $\hat S_{\theta}^{A}$. The 
results for this measurement have a macroscopic range and can tolerate 
increasing levels of (absolute) noise.

However it is crucial that 
the macroscopic nature of 
our 
result is clarified in the arrangement of Figure 1b. 
Here the field $\hat a_{-}¥$ is combined with field $\hat a_{+}¥$, to produce 
macroscopic fields $\hat a_{\pm}^{'}$, prior to the experimenter's   
selection of the angle $\theta$. These outgoing macroscopic fields 
$\hat a_{\pm}^{'}$ may then be regarded as the system at $A$. 
In this situation both fields $\hat a_{\pm}^{'}$ 
incident on the measurement apparatus, depicted by a  
polariser (or beam spitter) with the choice of $\theta$ in the 
Figure 1b, are macroscopic. (A similar 
description applies to the fields at $B$).

That a microscopic state was involved in the preparation of the 
spatially separated and propagating fields $\hat a_{\pm}^{'}$
 and $\hat b_{\pm}^{'}$ does not affect the macroscopic natre of our 
 work. it is important to realise that the measurement events at $A$ 
 and $B$ must be causally separated, and that by the time the fields 
 rach the measurement destinations $A$ and $B$, the original appartus 
 used to prepare the fields need no longer exist.

The important point is that this combining of fields 
which comes about as part of the 
state preparation can be clearly  
distinguished from amplification which comes after the selection of $\theta$, 
as part of the measurement process. This second-mentioned amplification comes 
about in all experiments, but does not imply that one can deduce 
``macroscopic elements of reality'' as we have defined it here. The 
``element of reality'' is a variable whose values refer 
to a physical quantity defined for a 
system, for example, the position of a particle. In the context of 
Einstein-Podolsky-Rosen and Bell arguments the system (for example the 
particle or photon field) has a well-defined 
meaning independent of the measuring apparatus (polariser or 
beamsplitter phase-shift combinations) and associated 
amplification. 
A macroscopic element of reality is a variable whose possible values 
are defined only with a macroscopic uncertainty. The value for the 
element of reality and its associated uncertainty have a clear 
meaning, and can be readily classified as macroscopic or 
not macroscopic. For example the uncertainty in the measured value for the 
position of a particle can be microscopic regardless of an amplified 
final readout value. In this work the ``element of reality'' refers to 
a photon number (actually a photon number difference) and ``macroscopic'' 
means a large photon number.

   \section{Conclusions}

  Our claim therefore is that earlier work $^{\cite{7,8}}$ 
suggestive of violations of local realism at a macroscopic level must be 
interpreted carefully before claiming a loss of local realism at a 
``macroscopic'' level. The failure of 
a Bell inequality in cases where the photon number can be macroscopic 
 but where 
measurement resolution is perfect may not automatically imply  
 the failure of a macroscopic local realism, as we have 
 defined it.

  In summary we have considered the concept of orders of local 
  realism, from macro- through meso- to microscopic, which apply to 
  experiments with increasing precision of measurement. Macroscopic local 
  realism excludes the possibility of macroscopic changes to a system $B$ 
  occurring as a result of events which occur simultaneously 
  at a spatially separated system $A$. This is as 
  opposed to local realism used in its 
  entirety, right down to the most microscopic level, which excludes all 
  orders of change. 
  
  We have derived Bell inequalities which, if violated 
  in experiments with limited resolution of photon number, will imply a 
  failure of these less restrictive forms of local realism.
   We claim that the proven failure, if ever achievable, of this 
 macroscopic local realism is conclusive evidence that 
 the ``startling'' properties apparently attributed to ``entangled  Schrodinger cat'' states 
 are inescapable. A class of quantum states (those showing a 
 violation of local realism for quadrature phase amplitudes) with this property has been proposed.

\begin{figure}
\includegraphics[scale=.85]{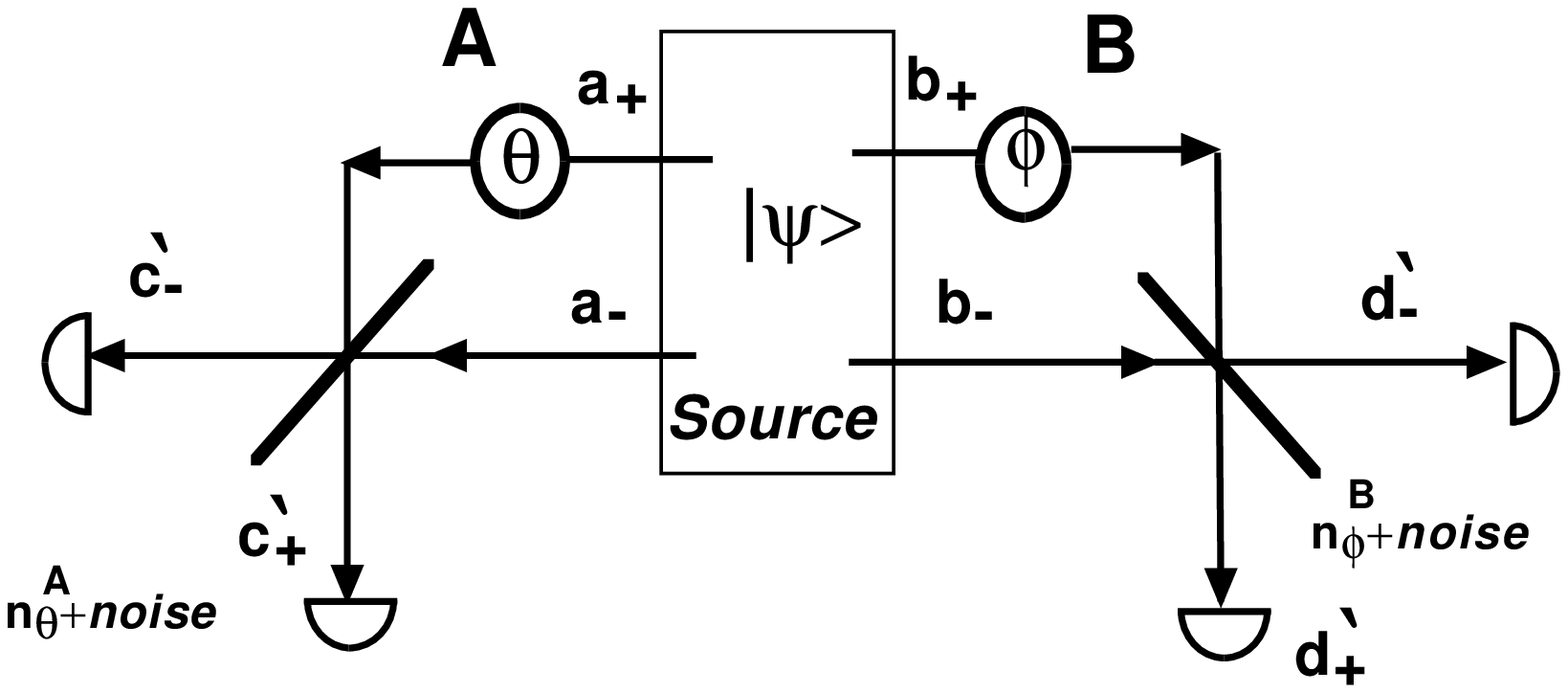}
\includegraphics[scale=.85]{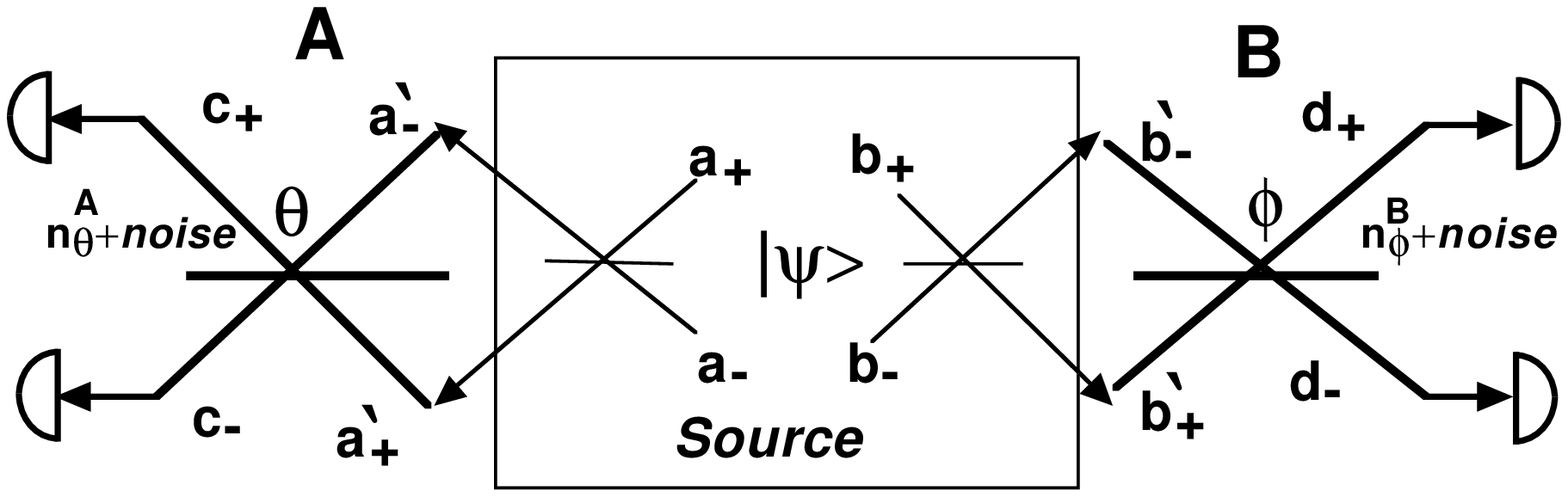}
 \caption[Schematic representation of a test of macroscopic local 
realism.]
{Schematic representation of our proposed test of macroscopic local realism. 
 (a) Measurement of spin operators $\hat S_{\theta}^A$ and $\hat S_{\phi}^B$. 
 This measurement scheme is equivalent to balanced homodyne 
 detection of the quadrature phase amplitudes $\hat X_{\theta}^A$ 
and $\hat X_{\phi}^B$ of the fields $\hat a_{-},\hat b_{-}$, in the limit of large 
 $\alpha$,$\beta$. In the proposed experiment $\hat a_{-},\hat b_{-}$ are of low 
 intensity while $\hat a_{+},\hat b_{+}$ are intense coherent-state 
 $|\alpha>$ ``local 
 oscillator'' fields. In this experiment 
 large intensities are 
 incident on each of the photodiode 
 detectors.
 (b) Importantly in this alternative arrangement 
 the fields $\hat a_{\pm}$ are first combined using a beam 
 splitter and phase shift so that both outgoing fields
  $\hat a_{\pm}^{'}$ incident on the measuring apparatus are 
  macroscopic. The measurement apparatus is 
  depicted here by the beam splitter with variable angle $\theta$, 
  although a polariser may also be possible for suitable states. A 
  similar arrangement occurs at $B$.
 In this experiment the entire boxed apparatus may be considered the source. 
 The measured quantity in terms of the $\hat a_{\pm},\hat b_{\pm}$ fields is 
 still $\hat S_{\theta}^A$ and $\hat S_{\phi}^B$ as above in (a).}%
\end{figure}

\begin{figure}
\includegraphics{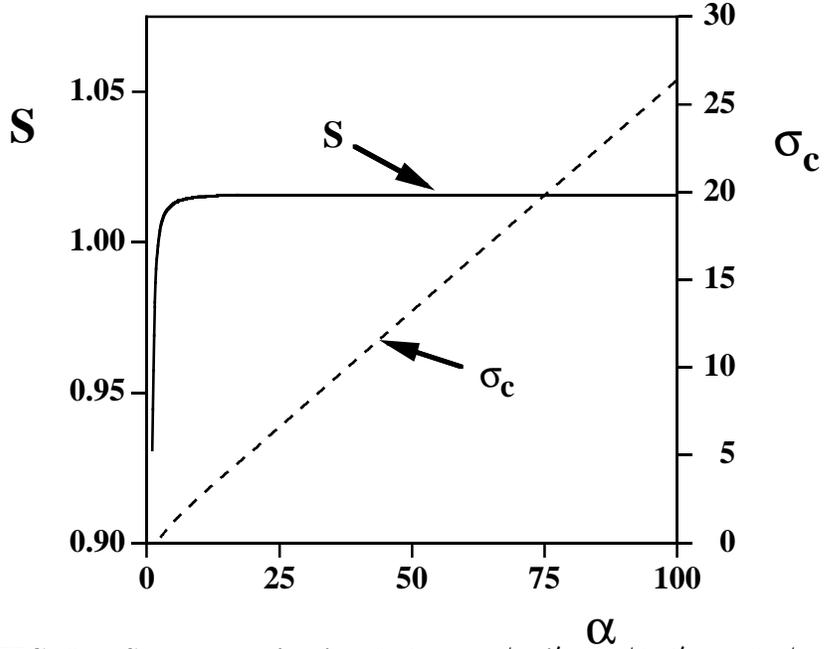}
\caption[Violations of the Bell's inequality found 
for the quantum state (16).]
{ $S$ versus $\alpha$, for 
$\theta=0,\phi=-\pi/4,\theta'=\pi/2,\phi'=-3\pi/4,\alpha=\beta$ for the 
quantum state (16) with no noise present. The dashed line gives the 
maximum noise $\sigma_{c}¥$ still giving a violation of the Bell 
inequality (7) for the above parameters, versus $\alpha$. Macroscopic 
values are possible with increasing $\alpha$.
}% 
\end{figure}

\begin{figure}
\includegraphics{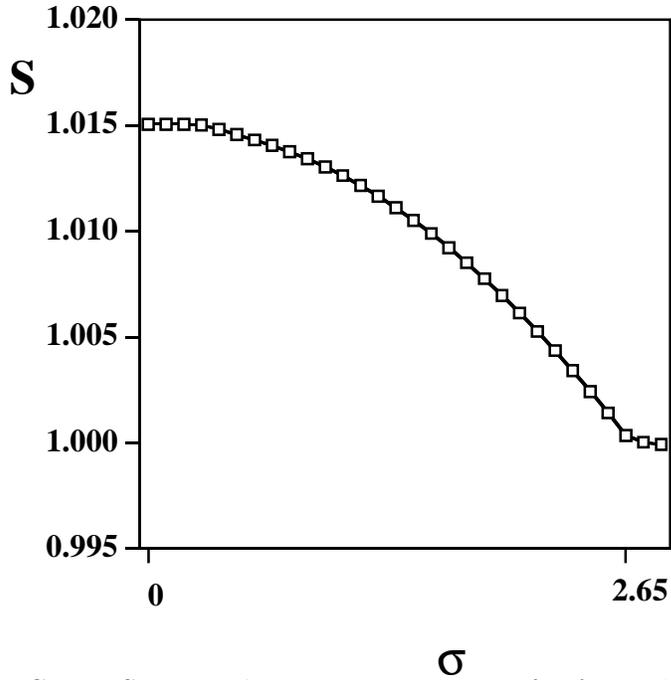}
\caption[Effect of noise on violations of the Bell's inequality found 
for the quantum state (16).]
{ $S$ versus the noise parameter $\sigma$, for 
$\theta=0,\phi=-\pi/4,\theta'=\pi/2,\phi'=-3\pi/4,\alpha=\beta$ for the 
quantum state (16), where $\alpha=10$.
}% 
\end{figure}

\begin{figure}
\includegraphics{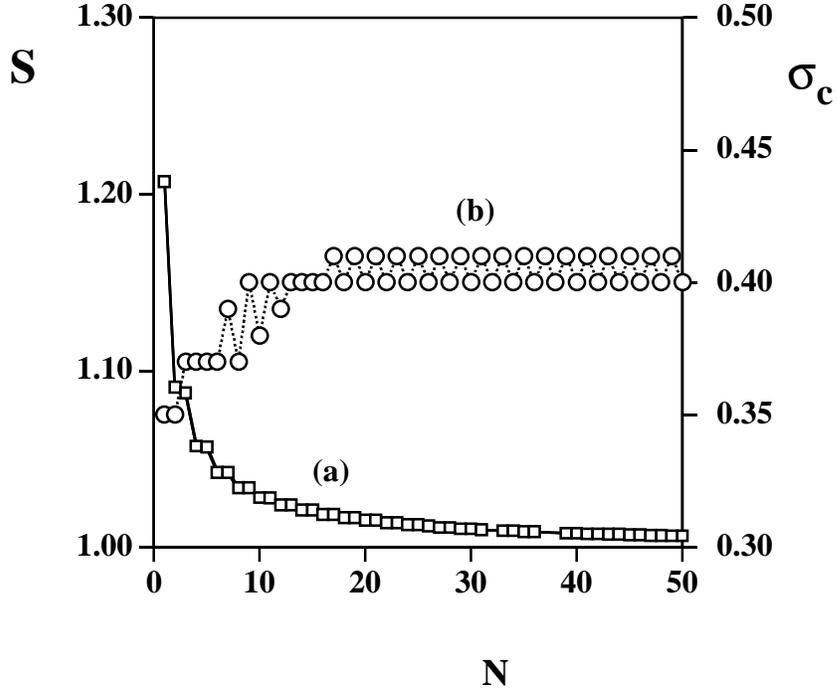}
\caption[Violations of the Bell's inequality found 
for the quantum state (22).]
{Line (a) gives $S$ versus $N$, for the 
quantum state (22) with no noise present.  Here we have selected 
the following relation between 
  the angles: $\phi -\theta=\theta' - \phi=\phi' -\theta'=\psi$ and 
  $\phi' -\theta=3\psi$ and optimised $S$ with respect to $\psi$.
Line (b) gives the maximum noise $\sigma_{c}$ still giving a violation of the Bell 
inequality (7) for the above parameters. In this case the  
cut-off noise $\sigma_{c}$ remains microscopic for large $N$.
}% 
\end{figure}

\end{document}